\documentclass[twocolumn,showpacs,preprintnumbers,prb]{revtex4-1}
\usepackage{dcolumn}
\usepackage{amssymb}
\usepackage{bm}
\usepackage{graphicx}
\usepackage{float}
\usepackage{subfigure}
\usepackage{stackengine}

\newcommand{\avg}[1]{\left< #1 \right>} 
\newcommand{\abs}[1]{| #1 |} 
\newcommand{\ket}[1]{| #1 \rangle} 
\newcommand{\bra}[1]{\langle #1 |} 

\begin{document}

\title{Sensitivity of the entanglement spectrum to boundary conditions as a characterization of the phase transition from delocalization to localization}
\author{Mohammad Pouranvari and Afshin Montakhab}
\affiliation{Department of Physics, College of Sciences, Shiraz
  University, Shiraz 71946-84795, Iran}

\date{\today}

\begin{abstract}
  Sensitivity of entanglement Hamiltonian spectrum to boundary
  conditions is considered as a phase detection parameter for 
  delocalized-localized phase transition. By employing one-dimensional
  models that undergo delocalized-localized phase transition, we study
  the shift in the entanglement energies and the shift in the entanglement
  entropy when we change boundary conditions from periodic to
  anti-periodic. Specifically, we show that both these quantities show
  a change of several orders of magnitude at the transition point in
  the models considered. Therefore, this shift can be used to indicate
  the phase transition points in the models.  We also show that both
  these quantities can be used to determine \emph{mobility edges}
  separating localized and delocalized states.
\end{abstract}

\maketitle

\section{Introduction}\label{Introduction}
Entanglement as a purely quantum phenomenon has been intensively
studied for the past decades.\cite{ref:Horodecki} It is thought to
underlie modern technologies such as quantum computing and
cryptography, to name a few.\cite{ref:chuang, ref:Ekert, ref:Steane,
  ref:Gisin} Recently, entanglement has been used intensively to study
condensed-matter systems as well. Entanglement is a measure of how
much quantum correlation exists in multipartite quantum system. In
condensed-matter, systems that exhibit continuous phase transition are
marked by a critical point, where the system becomes highly correlated
with power-law (long-range) correlations. It is therefore not
surprising that entanglement can be used as a parameter to
characterize phase transition and critical points in quantum many-body
system,\cite{ref:osterloh, ref:amico, ref:haque, ref:kitaev} although
there exist some debates.\cite{ref:chandran, ref:min-fongyang}

There are several measures of entanglement\cite{ref:vedral} by
which various authors have characterized different phases and
phase transitions.\cite{ref:vidal,ref:osborne,ref:shi, ref:vidal2}
von Neumann entanglement entropy (EE), as the most popular and
standard measure of entanglement in a pure state, has been
frequently used. In a bipartite approach, one can partition the
system in various ways, as in momentum space, \cite{ref:mondragon,
ref:andrade} or a combination of momentum and orbital partition,
\cite{ref:mondragon2} or various other choices.
\cite{ref:othercut}  In addition,  other authors have advocated a
multipartite approach where entanglement finds a more (extensive)
thermodynamic interpretation. \cite{ref:wei,
ref:love,
  ref:chhajlany, ref:afshin1, ref:afshin2} People also use spectrum of the reduced density matrix
\cite{ref:LiHaldane} to distinguish different phases. In addition,
it is also shown that eigen-modes of the entanglement Hamiltonian
may carry some useful physics.\cite{ref:pouranvariyang1,
ref:pouranvariyang2,
  ref:pouranvariyang5}

Among the various phase transitions in condensed-matter physics,
Anderson phase transition between a localized and an extended
(delocalized) phase is of particular interest. Various authors
have also studied such a transition in the light of entanglement.
For example, in Ref. [\onlinecite{ref:berkovitsprl2015}] the
probability distribution of the EE is used to characterize
different phases in a one-dimensional wire with attractive
interaction. In Ref. [\onlinecite{ref:chakravarty}] it is shown
that EE is non-analytic at the delocalized-localized phase
transition point. A finite-size scaling of the EE is done in Ref.
[\onlinecite{ref:zhao}] to characterize the Anderson transition
and to  obtain the critical exponents. The dependence of the EE upon
mean free path in a free fermion model and upon the localization
length in interacting model with Anderson transition is studied in
Ref. [\onlinecite{ref:pouranvariyang4}] and Ref.
[\onlinecite{ref:berkovitsprl2012}], respectively. More recently,
the role of the entanglement in interacting models and its
relation to thermalization has been emphasized.
\cite{ref:bardarson, ref:nag, ref:filho, ref:geraedts,
  ref:khemani, ref:yang, ref:berkovits, ref:vidmar}

In this paper, we intend to study localization-delocalization
phase transition by introducing another related quantity as a
phase detection parameter, namely the sensitivity of the
entanglement energies to boundary condition. Effect of the
(sub)system boundary condition on the entanglement properties has
been studied before. Here we mention some of these studies: in
Ref. [\onlinecite{ref:laflorencie2006}] the effect of the open
boundary condition (in contrast to periodic boundary condition) on
the entanglement entropy is calculated. Effect of an impurity located on
the subsystem boundary is studied for a Luttinger liquid in Ref.
[\onlinecite{ref:levine}]. In Ref. [\onlinecite{ref:peschel2005}]
defects at the boundary of the subsystem for a tight binding model
is considered as impurities on the hopping elements and on-site
energies, and their effects on the entanglement spectrum and
central charge. Our approach is different from above in a sense
that we consider the effect of boundary conditions in different
phases, and use this as a detection mechanism for the phase
transition.

Regarding delocalized-localized transition, there are several
methods to distinguish different phases. The most widely used
method is the statistics of the level
spacing,\cite{ref:Shklovskii} which we only need the
eigen-energies rather than the eigen-states. In Ref.
[\onlinecite{ref:edward}], Edwards and Thouless study the
sensitivity of the eigen-energies of a system's Hamiltonian to the
boundary conditions. When boundary conditions change from periodic
boundary condition (PBC) to anti-periodic boundary condition
(APBC) the shift in the eigen-energies is used to distinguish
localized and de-localized phases. The basic idea is the
following: if the eigen-mode is localized, it does not ``see'' the
boundaries and thus it is not affected by any change in the
boundary conditions and the corresponding eigen-energy does not
alter. On the other hand, when the eigen-mode is extended, it is
affected by what happens at the boundary; the change in the
corresponding eigen-energy being comparable to the spacing between
eigen-energies. They used the amount of this shift as a criterion
for detecting the Anderson phase transition. This shift in the
eigen-energies is related to the transmission and subsequently to
the conductance of the system.\cite{ref:anderson, ref:economou,
ref:kramer}

On the other hand, there are similarities between the eigen-modes of
the Hamiltonian and the eigen-modes of the entanglement Hamiltonian,
specially between the eigen-modes of the Hamiltonian at the Fermi
level $\ket{E_F}$, and the maximally entangled mode
$\ket{MEM}$.\cite{ref:footenote} In Ref.
[\onlinecite{ref:pouranvariyang2}] we demonstrated this similarity by
employing two one-dimensional free fermion models that exhibit
localized-delocalized (LD) phase transition. We found that both
$\ket{E_F}$ and $\ket{MEM}$ are extended in the delocalized phase and
both are localized in the localized phase. Also, their overlap is
substantial in the delocalized phase or at least at the phase
transition point. In short, eigen-modes of the entanglement
Hamiltonian and specially the $\ket{MEM}$ carry on some physics of the
$\ket{E_F}$.  In this paper, we further address this similarity by
showing that one can extract localization properties of the system by
studying \emph{entanglement Hamiltonian} instead of Hamiltonian of the
system. We conjecture that if the entanglement Hamiltonian eigen-mode
is extended the corresponding entanglement energy is affected by the
boundary conditions and if it is localized, then the corresponding
entanglement energy does not change.

Accordingly, in specific one-dimensional free fermion models that
undergo LD phase transition, we change the boundary condition from
PBC to APBC and see how the entanglement energies and thus
entanglement entropy changes. We show numerically that the shift
in the entanglement Hamiltonian spectrum is considerable in the
delocalized phase but it is negligible in the localized phase.
Thus, it can be used as a characterization of LD phase transition.
Furthermore, we also show that the same ideas can be used to
identify mobility edges. We would like to mention that the
one-dimensional models we consider here have theoretical
relevance, but have also recently been found to have experimental
relevance as well. \cite{ref:diener, ref:lahini}

The remainder of the paper is as follows: in Sec. \ref{model}, we
explain the method for calculating the entanglement spectrum and
entanglement entropy. We also explain the one-dimensional models that
are used in this paper to verify our conjecture. In Sec.
\ref{results}, we present the main result of this paper: we study the
effect of the change in the boundary conditions on entanglement
Hamiltonian spectrum, and entanglement entropy. We also show that the
shift, \emph{only} in the smallest magnitude entanglement energy, is
enough to characterize the phase transition. As an extra check, for a
one-dimensional model with mobility edges, we show that the shift
locates the mobility edges for the whole spectrum. We close in
Sec. \ref{conclusion} with a summary and concluding remarks.

\section{Methods and Models}\label{model}
If a system is in a pure state $\ket{\Psi}$, density matrix will be
$\rho= \ket{\Psi} \bra{\Psi}$. We divide the system into two
subsystems $A$ and $B$ and for each subsystem the reduced density
matrix is obtained by tracing over degrees of freedom of the other
subsystem: $\rho^{A/B}=tr_{B/A} (\rho)$. Block von Neumann
entanglement entropy between the two subsystems is
$EE=-tr(\rho^{A}\ln{\rho^{A}})=-tr(\rho^{B}\ln{\rho^{B}})$. For a
single Slater-determinant ground state, the reduced density matrix of
each subsystem can be written as:
\begin{equation} \label{rho}
  \rho^{A/B}=\frac{1}{Z} e^{-H^{A/B}},
\end{equation}
where $H^{A/B}$ is the free-fermion \emph{entanglement} Hamiltonians
($Z$ is determined by $tr \rho^{A/B}=1$):
\begin{equation}
H^{A/B} = \sum_{ij} h_{ij}^{A/B}  c_{i} ^{\dagger} c_{j},
\end{equation}
where $ c_{i} ^{\dagger} (c_{i})$ is the fermionic creation (annihilation)
operator for site $i$.

To calculate entanglement energies $\epsilon$'s, i.e. the eigen-values
of the $h^{A/B}$ matrix, we use method of Ref. [\onlinecite{correl}]:
we divide the system in two parts, subsystem $A$ from site $1$ to
$N_A$ and the rest as subsystem $B$. We diagonalize the correlation
matrix of a subsystem, say $A$
\begin{equation}
C_{i,j}=\avg{c_{i} ^{\dagger} c_{j}},
\end{equation}
(where $i$ and $j$ go from $1$ to $N_A$) and find its eigen-values $\{ \zeta \}$. Eigen-values of
the correlation matrix and those of the entanglement Hamiltonian are related
as:
\begin{equation}
\zeta_i=\frac{1}{1+e^{\epsilon_i}},
\end{equation}
and EE will be given as:
\begin{equation}\label{EE}
\text{EE}=-\sum_{i=1} ^{N_A} [\zeta_i \ln(\zeta_i)+(1-\zeta_i) \ln(1-\zeta_i)],
\end{equation}

Next, we introduce lattice models we work with in this paper. They are
one-dimensional free fermion tight binding models with constant
nearest neighbor coupling $t$ and on-site energies $\phi_n$:

\begin{equation}\label{oh}
  H=t\sum_{n=1}^{N} (c_n^{\dagger} c_{n+1}+c_{n+1}^{\dagger} c_n)+
  \sum_{n=1}^{N} \phi_n c_n^{\dagger} c_n,
\end{equation}

The first model is random dimer model (RD)\cite{dunalp} where
$\phi_n$'s are randomly chosen from one of two independent on site
energies $\phi_a$ or $\phi_b$. One of the site energies (we choose it
to be $\phi_b$) is assigned to neighboring pairs of lattice sites. As
shown by Dunalp \emph {et al.}\cite{dunalp}, when
$-2t \leq \phi_a-\phi_b \leq 2t$, states at the resonant energy,
$E_{res}=\phi_b$, are delocalized. Here we set $t=-1$ and $\phi_a=0$,
thus when $-2 \leq \phi_b \leq 2$ system is delocalized at the
resonant energy $E_F=E_{res}=\phi_b$, and is localized otherwise.

Another model with the Hamiltonian of the form Eq. (\ref{oh}) has
on-site energies:
\begin{equation}\label {gAA}
  \phi_n = 2 \lambda \frac{\cos{(2\pi n b)}}{1-\alpha\cos{(2\pi n b)}},
\end{equation}
where $b = \frac{1+\sqrt{5}}{2}$ is the golden ratio, so that it
has incommensurate period with respect to lattice period (we set
the lattice constant to be $1$). This model is neither completely
periodic (with extended eigen-modes) nor completely random (with
localized eigen-modes) and, as illustrated in Ref.
[\onlinecite{ref:dassarma}], it has \emph{mobility edges}
separating localized and delocalized states at:
\begin{equation}\label{me}
   E_{\text{mobility edge}} =\frac{ 2 sgn(\lambda)(\abs{t}-\abs{\lambda})}{\alpha},
\end{equation}
where in our calculation we set $t=-1$.

An special case of Eq. (\ref{gAA}) with $\alpha=0$ is the
Aubry-Andre (AA) model\cite{ref:aubryandre} which has a phase
transition at $\lambda=1$. All eigen-states for $\lambda < 1$ are
delocalized whereas they become all localized for $\lambda
> 1$. Thus the AA model does not have mobility edges.

\section{Sensitivity of Entanglement Properties to Boundary Condition}
\label{results}

\begin{figure}
  \centering
  \begin{subfigure}{}%
    \includegraphics[width=0.235\textwidth]{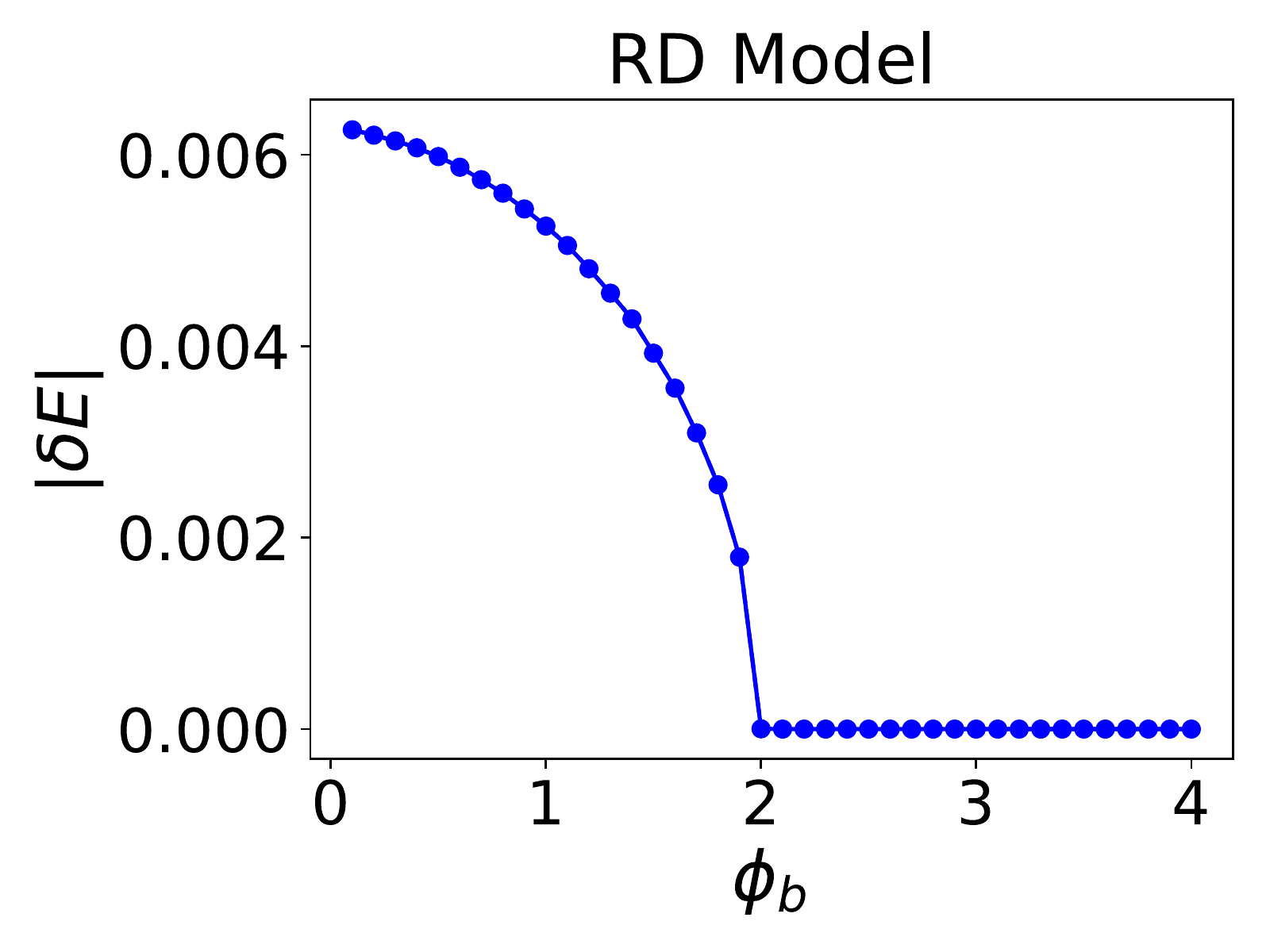}
  \end{subfigure}%
  ~%
  \begin{subfigure}{}%
    \includegraphics[width=0.235\textwidth]{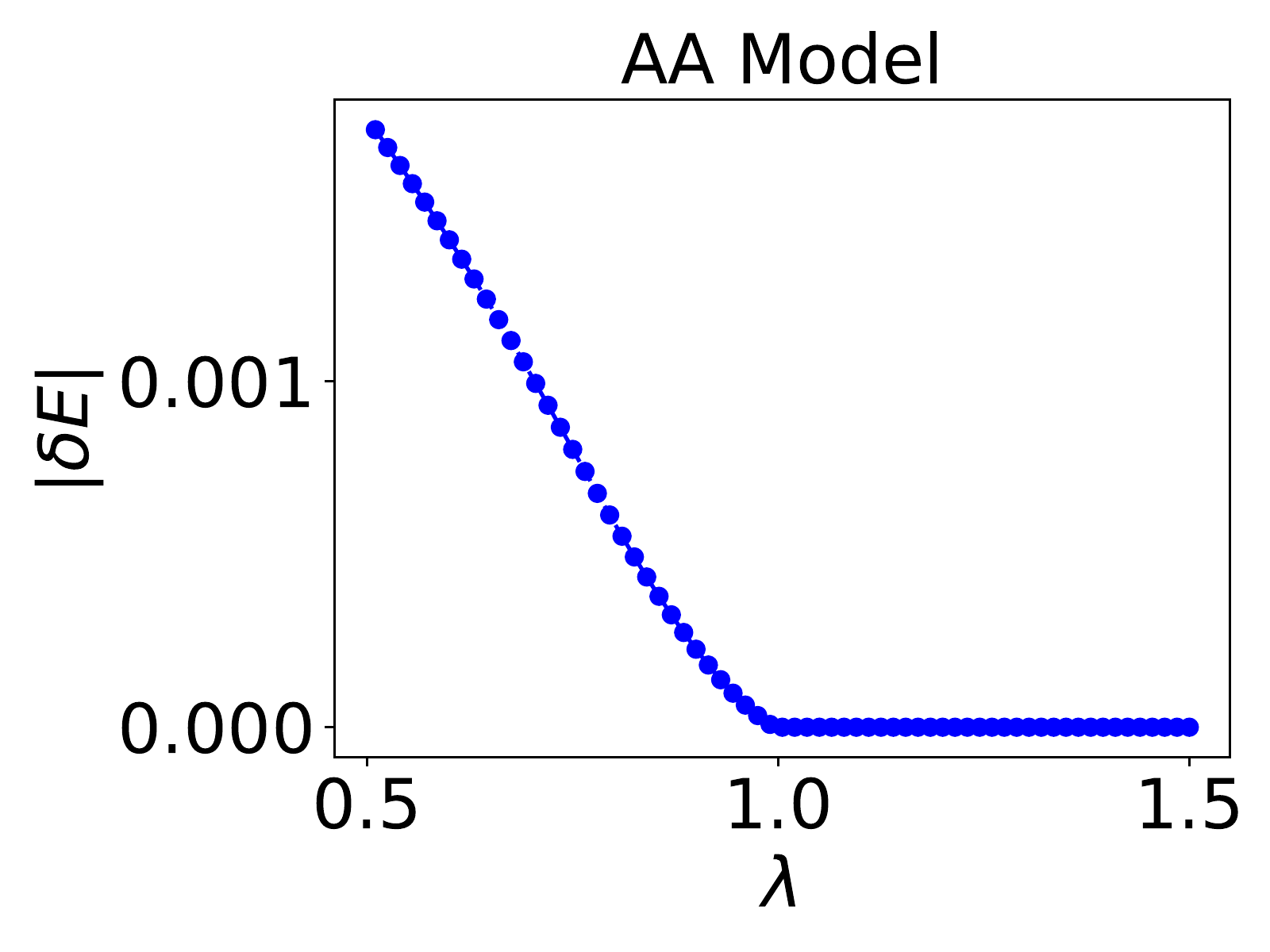}
  \end{subfigure}
  \caption{Left panel: disorder average of magnitude of shift in the
    Hamiltonian eigen-energy at the Fermi level for RD model when we
    change boundary condition from PBC to APBC as a function of
    $\phi_b$. System size $N=1000$, disorder average is over $1000$
    samples. Right panel: magnitude of shift in the Hamiltonian
    eigen-energy at the Fermi level for AA model as a function of
    $\lambda$. System size $N=2000$. \label{fig:dE_RDM_AA}}
\end{figure}

As mentioned above, Edward and Thouless\cite{ref:edward} used the
sensitivity of the eigen-energies of the system's Hamiltonian to
boundary conditions to distinguish localized from delocalized
phases. They used the geometrical average of shifts in the whole
spectrum. For comparison with our method, we calculate the magnitude
of shift of the eigen-energy at the Fermi level $|\delta E|$ when we
change the boundary conditions. In our one-dimensional models we apply
PBC by imposing $\psi_{N+1}=+\psi_{1}$ and APBC by
$\psi_{N+1}=-\psi_{1}$. The results are plotted in Fig.
\ref{fig:dE_RDM_AA}. For both RD and AA models $|\delta E|$ in the
delocalized phase is non-zero, it gradually becomes smaller as we
approach the phase transition point, and in the localized phase it
becomes zero. As seen in Fig. \ref{fig:dE_RDM_AA}, $|\delta E|$
behaves much as an order parameter in standard phase transition.

\subsection{Shift in the Entanglement Hamiltonian Spectrum and
  Entanglement Entropy}

\begin{figure*}
  \centering
  \begin{subfigure}{}%
    \def\big{\includegraphics[trim={10pt 0pt 10pt 10pt},clip,width=0.4\textwidth]{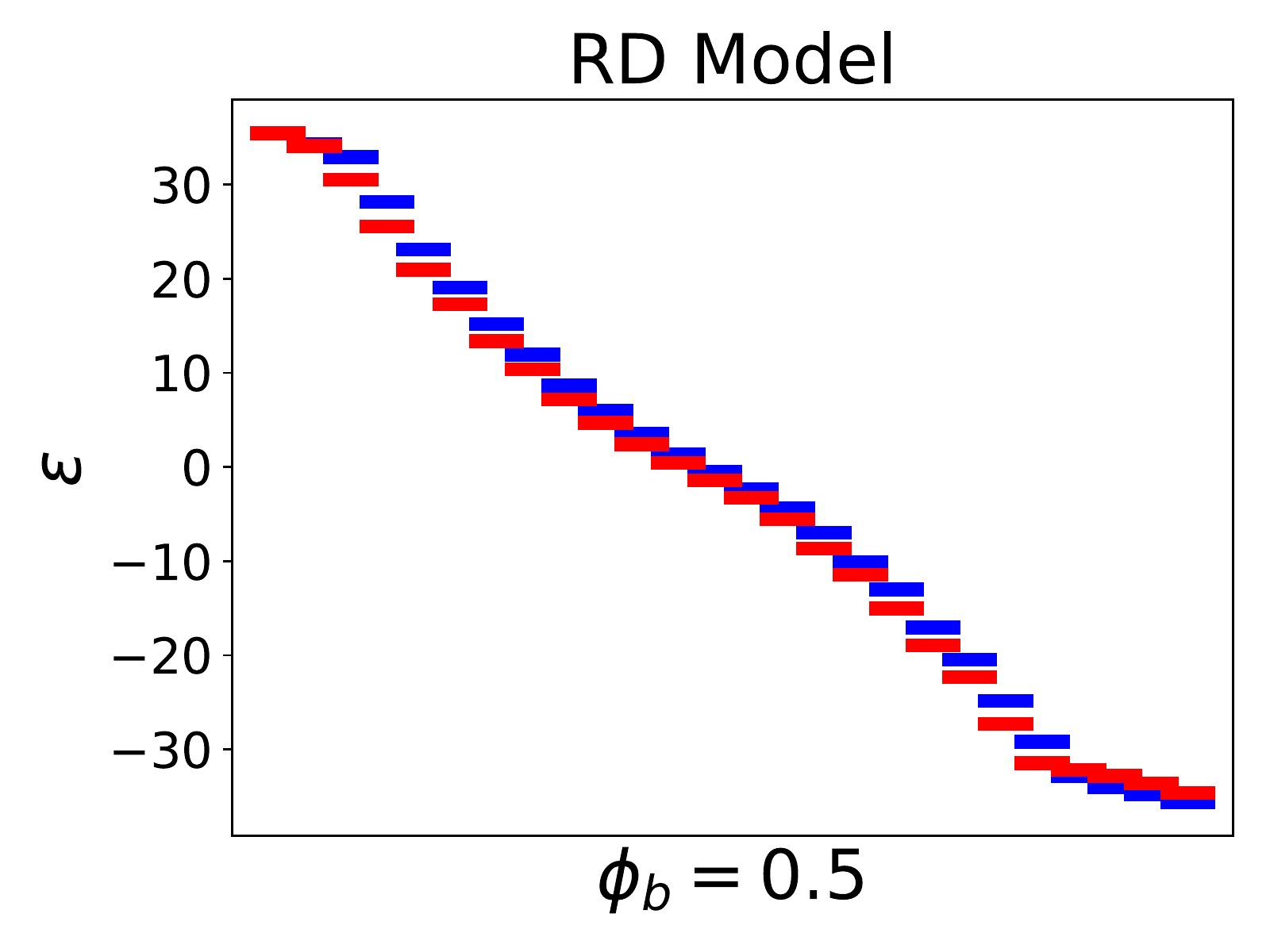}}
    \def\little{\includegraphics[trim={5pt 0pt 10pt 10pt},clip,width=0.14\textwidth]{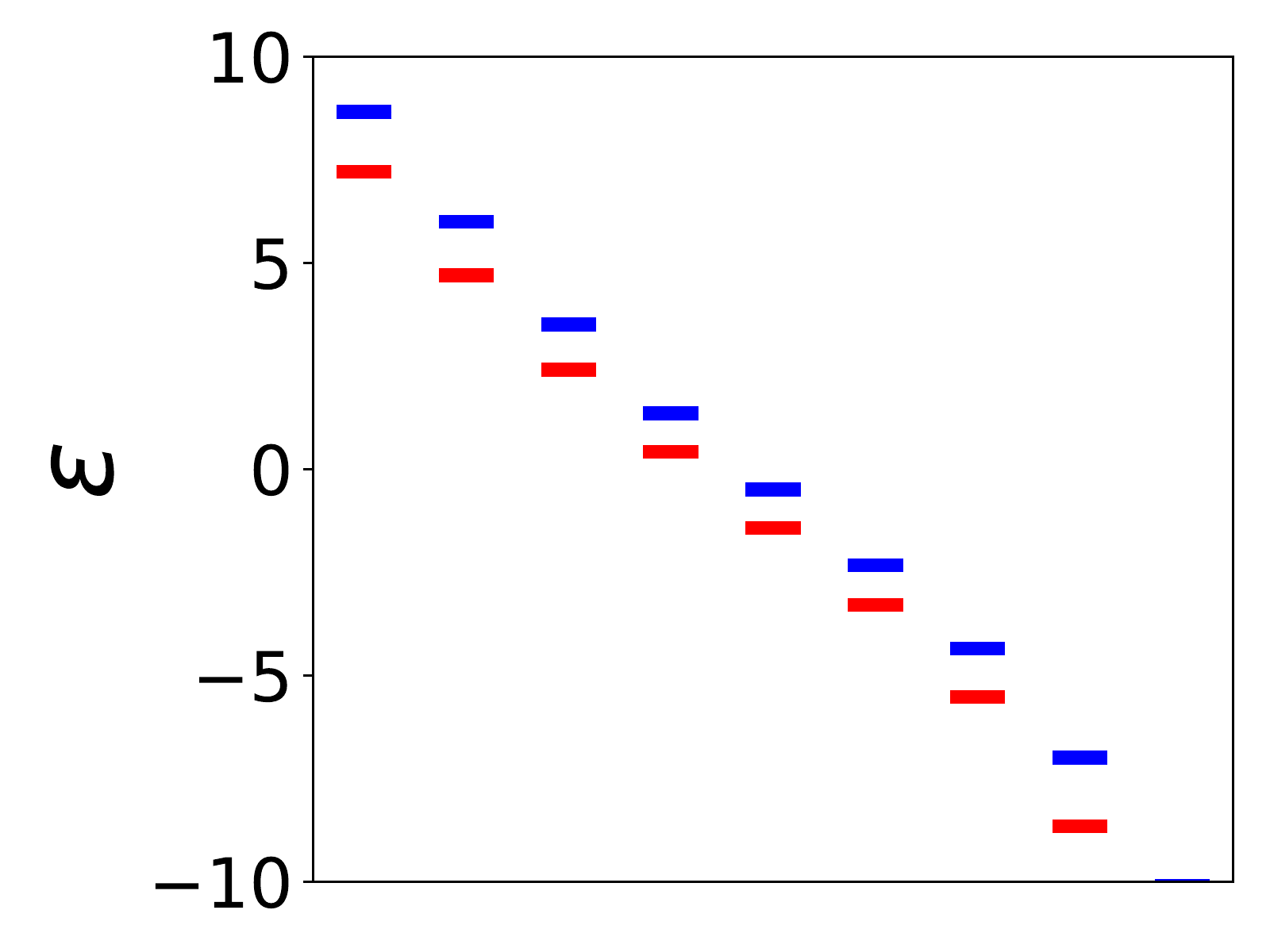}}
    \stackinset{r}{16pt}{t}{16pt}{\little}{\big}
  \end{subfigure}%
  ~%
  \begin{subfigure}{}%
    \def\big{\includegraphics[trim={10pt 0pt 10pt 10pt},clip,width=0.4\textwidth]{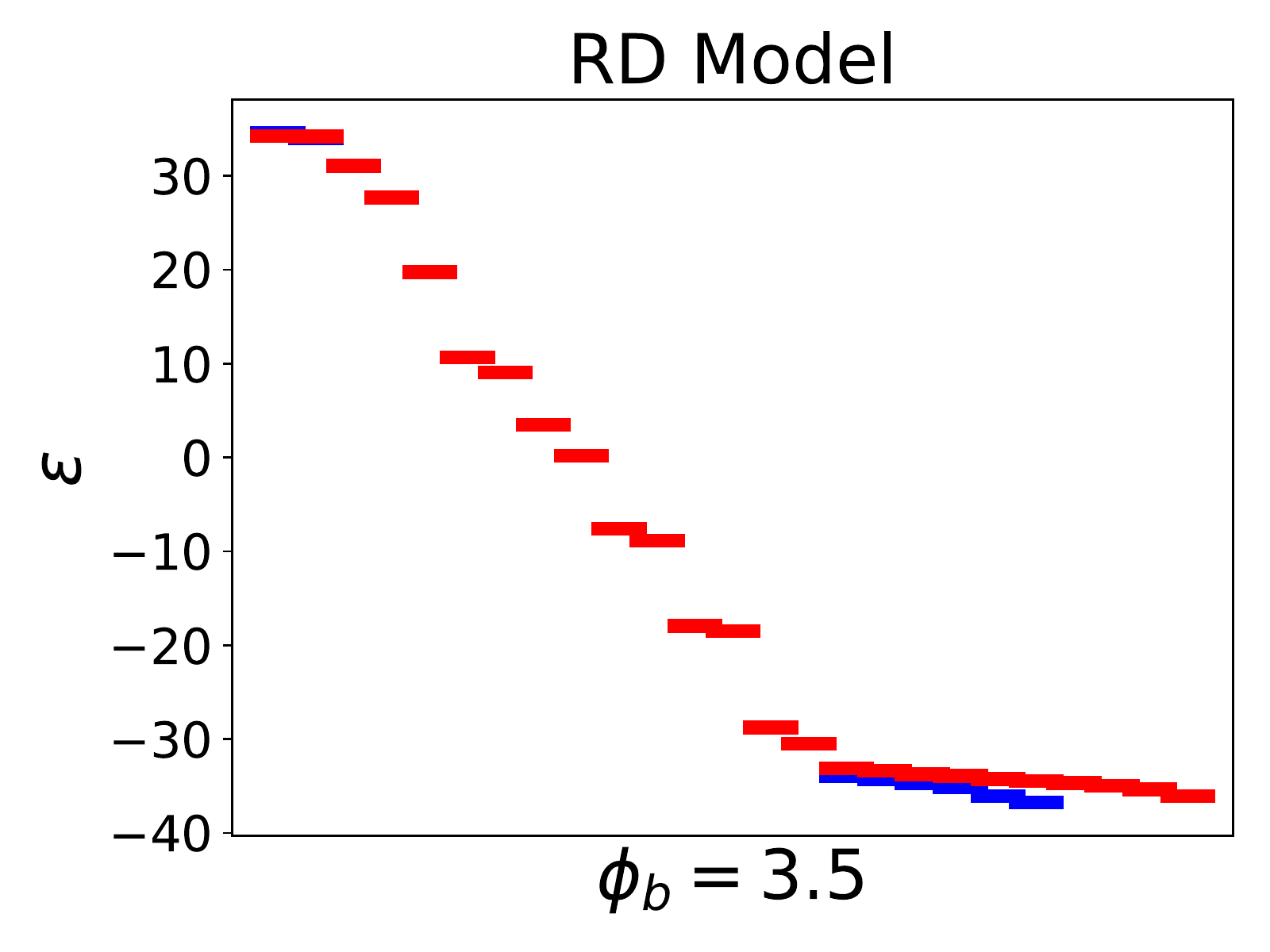}}
    \def\little{\includegraphics[trim={10pt 0pt 10pt 10pt},clip,width=0.14\textwidth]{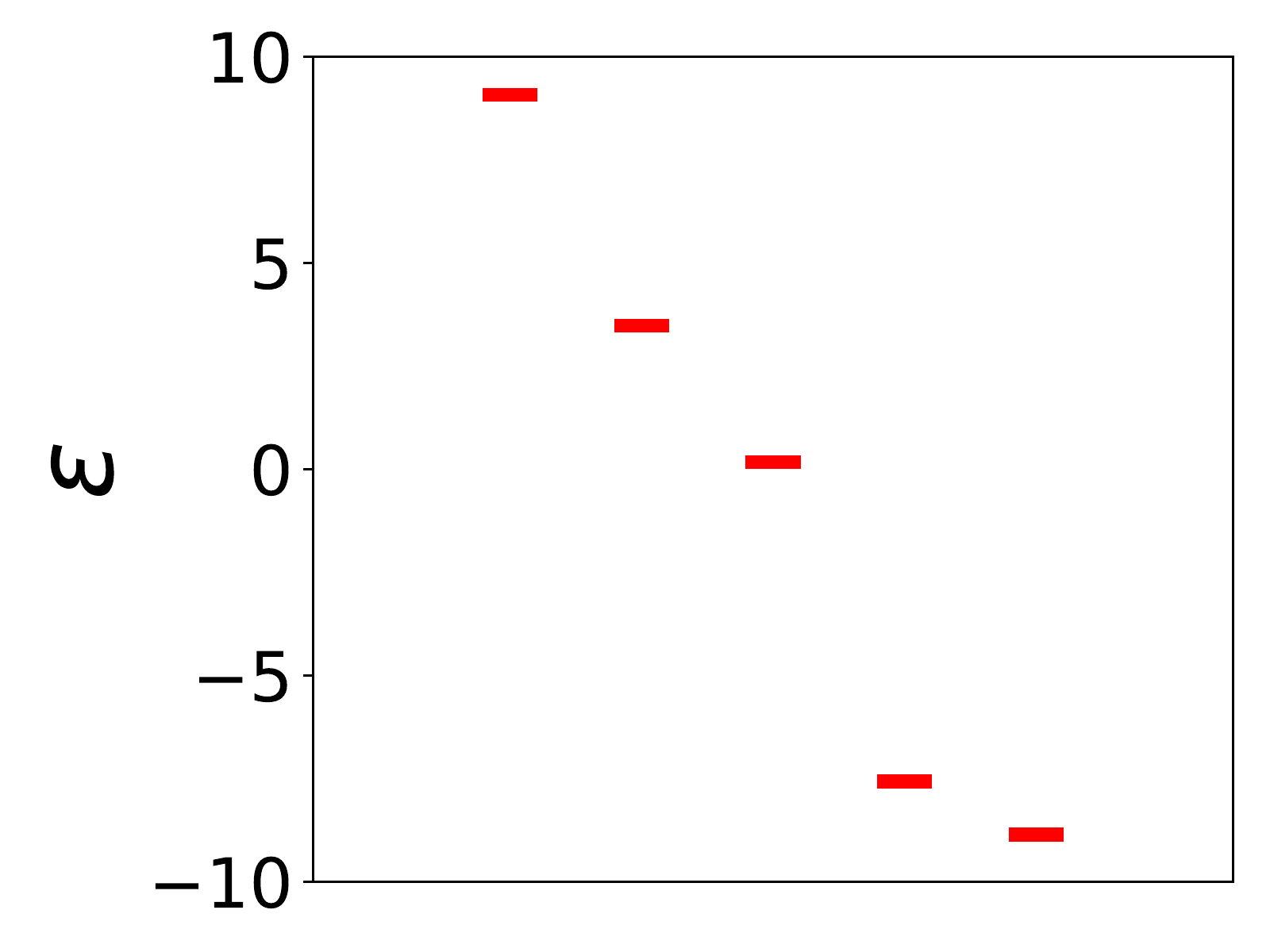}}
    \stackinset{r}{16pt}{t}{16pt}{\little}{\big}
  \end{subfigure}
  \caption{(color online) eigen-values of entanglement
    Hamiltonian of RD model for $\phi_b=0.5$ (in delocalized phase)
    and for $\phi_b=3.5$ (in localized phase). Spectrum with PBC is
    plotted in blue, and with APBC in red. Inset plots are zoomed
    plots to show a few eigen-values close to zero. At each $\phi_b$
    only one sample is considered without taking disorder
    average. We choose $N=60, N_A=30$. \label{fig:spectrumRDM}}
\end{figure*}

To study the sensitivity of the entanglement to the boundary
conditions, let us first examine the \emph{spectrum} of the
entanglement Hamiltonian $\{\epsilon\}$ of one subsystem (here we
choose subsystem $A$) when we change boundary condition from PBC to
APBC. In Fig. \ref{fig:spectrumRDM}, we plot spectrum of the
entanglement Hamiltonian in RD model for both cases of PBC and APBC at
two different $\phi_b$'s. We choose a $\phi_b=0.5$ in the delocalized
phase and a $\phi_b=3.5$ in the localized phase. Only one sample is
considered at each $\phi_b$. There is a shift between two spectra in
the delocalized phase, whereas they are the same in the localized
phase --- not for the whole spectrum but at least for those
$\epsilon$'s close to zero, which are more important since they have
more contributions to the entanglement entropy EE, Eq. (\ref{EE}).

\begin{figure*}
  \centering
  \begin{subfigure}{}%
    \includegraphics[width=0.45\textwidth]{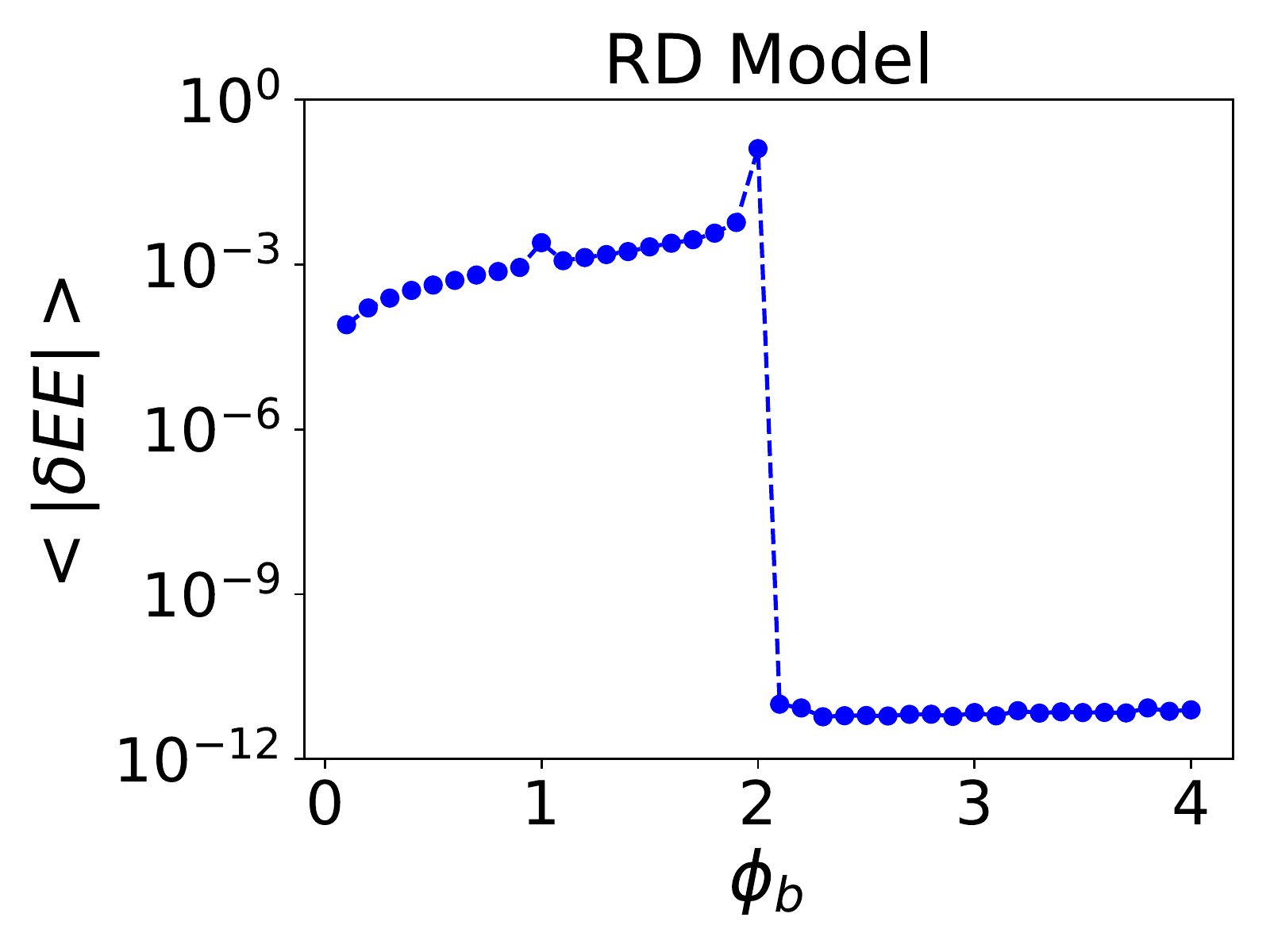}
  \end{subfigure}%
  ~%
  \begin{subfigure}{}%
    \includegraphics[width=0.45\textwidth]{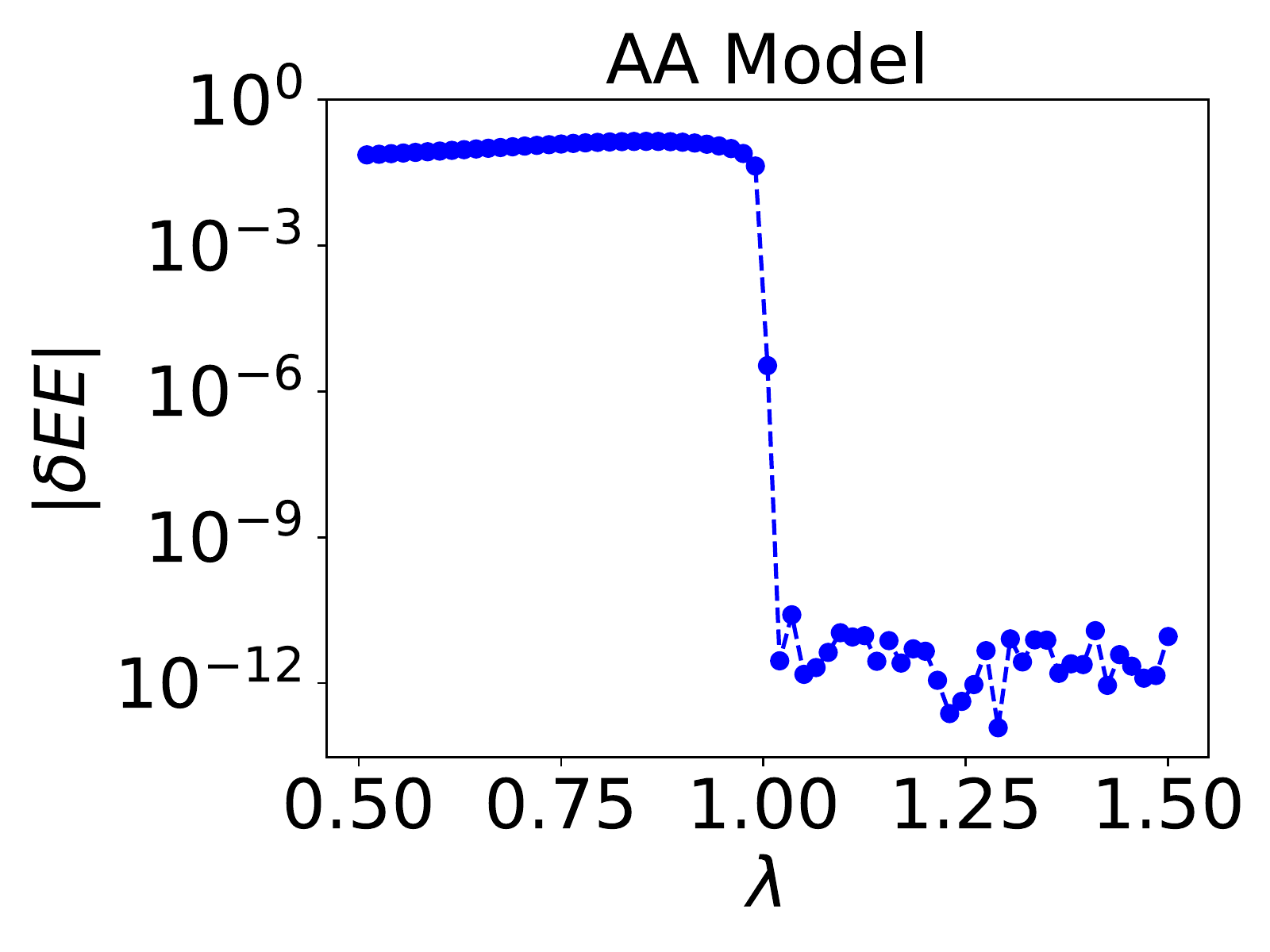}
  \end{subfigure}
  \caption{Left panel: disorder average of magnitude of shift in the
    entanglement entropy $<|\delta EE|>$ when we change boundary
    condition from PBC to APBC for RD model. In the delocalized phase
    EE $\sim$ 2.5 and in the localized phase EE $\sim$ 0.5. $N=1000$. Right
    panel: magnitude of shift in EE $|\delta EE|$ for AA model. In the
    delocalized phase EE $\sim$ 2.2 and in the localized phase EE
    $\sim$ 0.4. $N=2000$. \label{fig:dEE}}
\end{figure*}

To see the shift in the spectrum more quantitatively, we calculate the
magnitude of
shift in the \emph{entanglement entropy} $|\delta EE|$ for both RD and
AA models, when we change boundary condition (see
Fig. \ref{fig:dEE}). Since in the delocalized phase, the spectrum is
modified, the change in the entanglement entropy is non-zero, although
very small compare to the EE value at each point. But, in the
localized phase the change is much smaller and very close to zero.

\subsection{Shift in the Smallest Magnitude Entanglement Energy}
\begin{figure*}
  \begin{subfigure}{}%
    \centering
    \def\big{\includegraphics[trim={8pt 0pt 10pt 10pt},clip,width=0.4\textwidth]{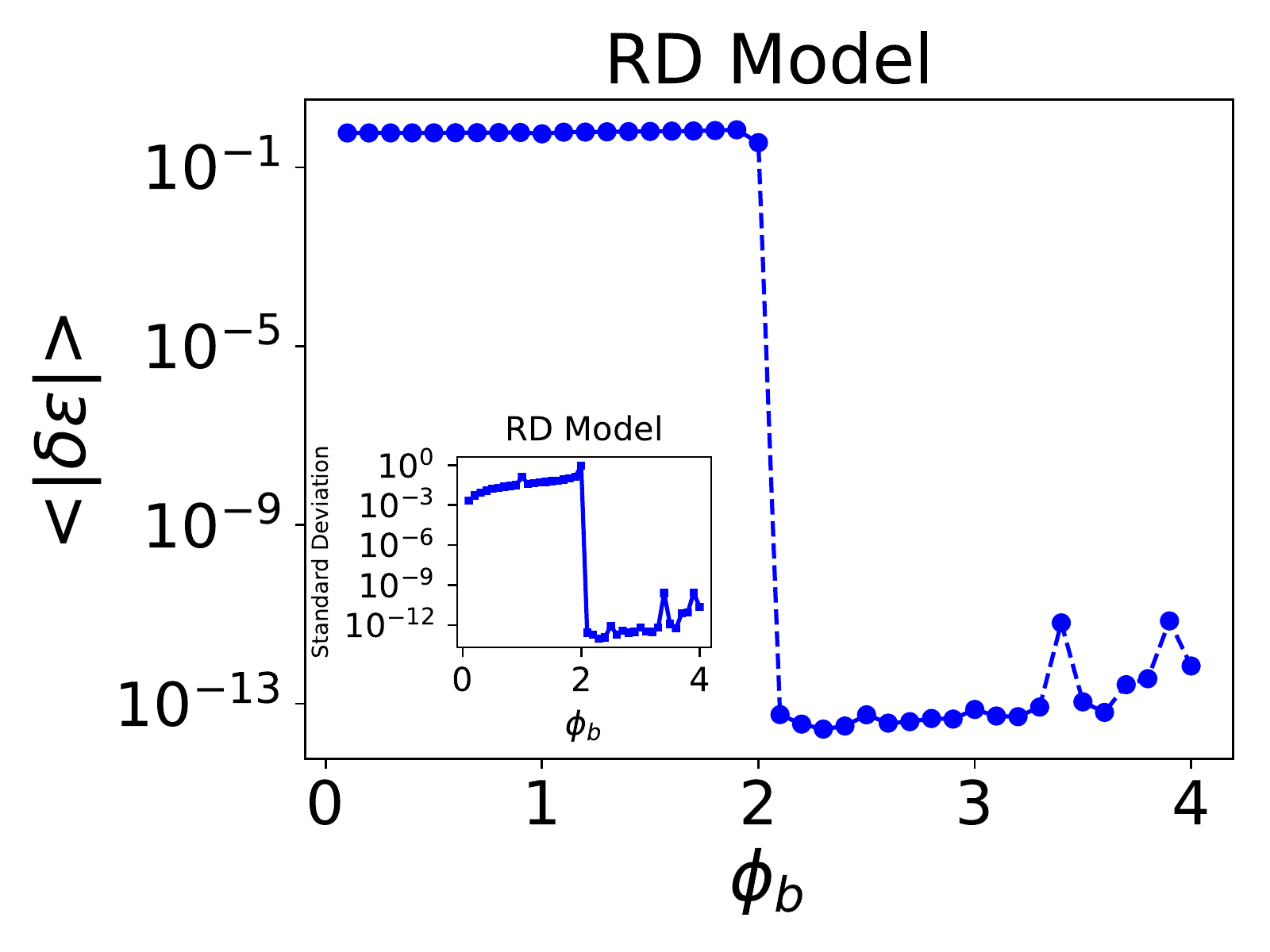}}
    \def\little{\includegraphics[trim={8pt 0pt 10pt 10pt},clip,width=0.14\textwidth]{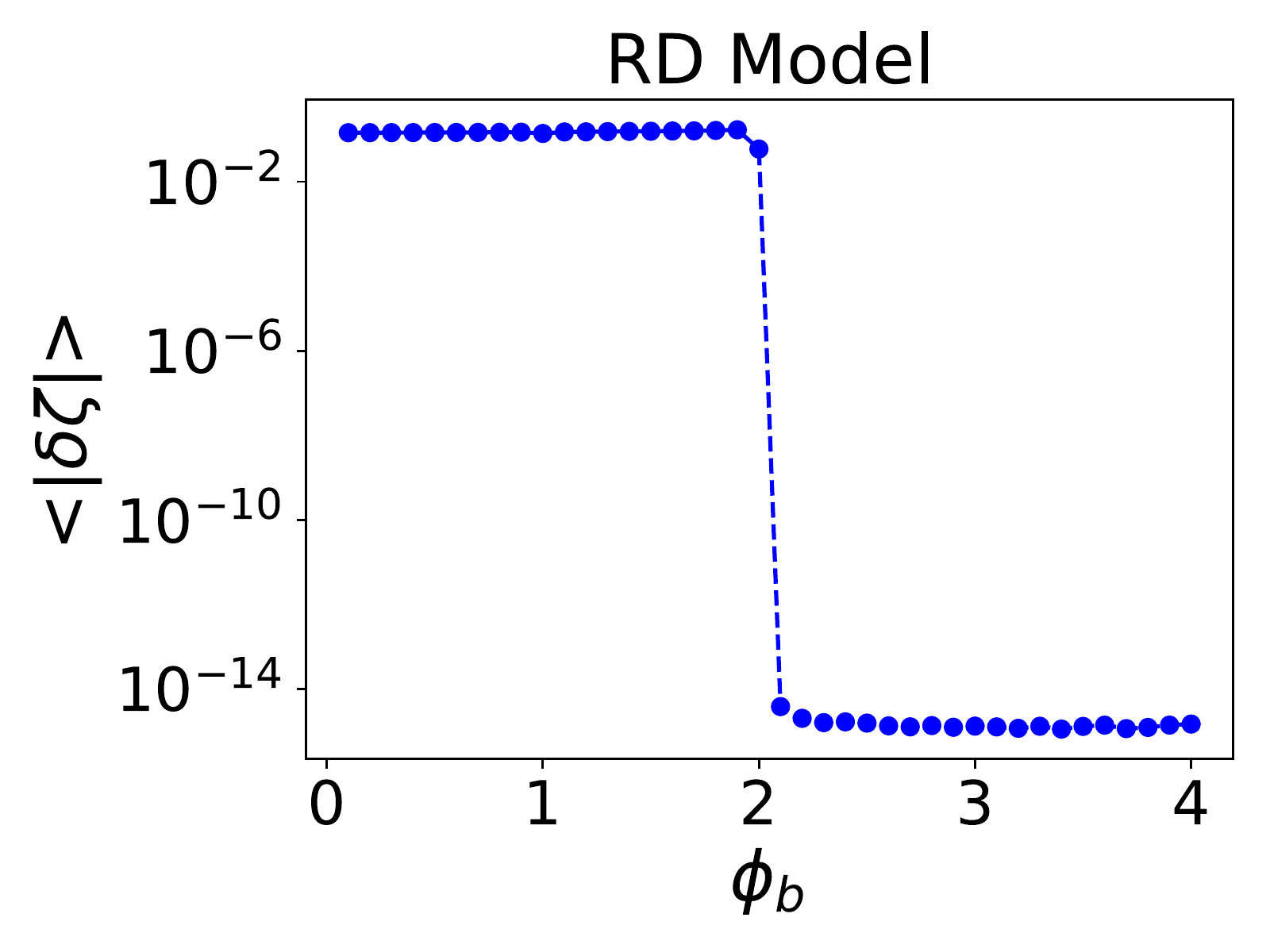}}
    \stackinset{r}{7pt}{t}{15pt}{\little}{\big}
  \end{subfigure}%
  ~%
  \begin{subfigure}{}%
    \centering
    \def\big{\includegraphics[trim={10pt 0pt 10pt 10pt},clip,width=0.4\textwidth]{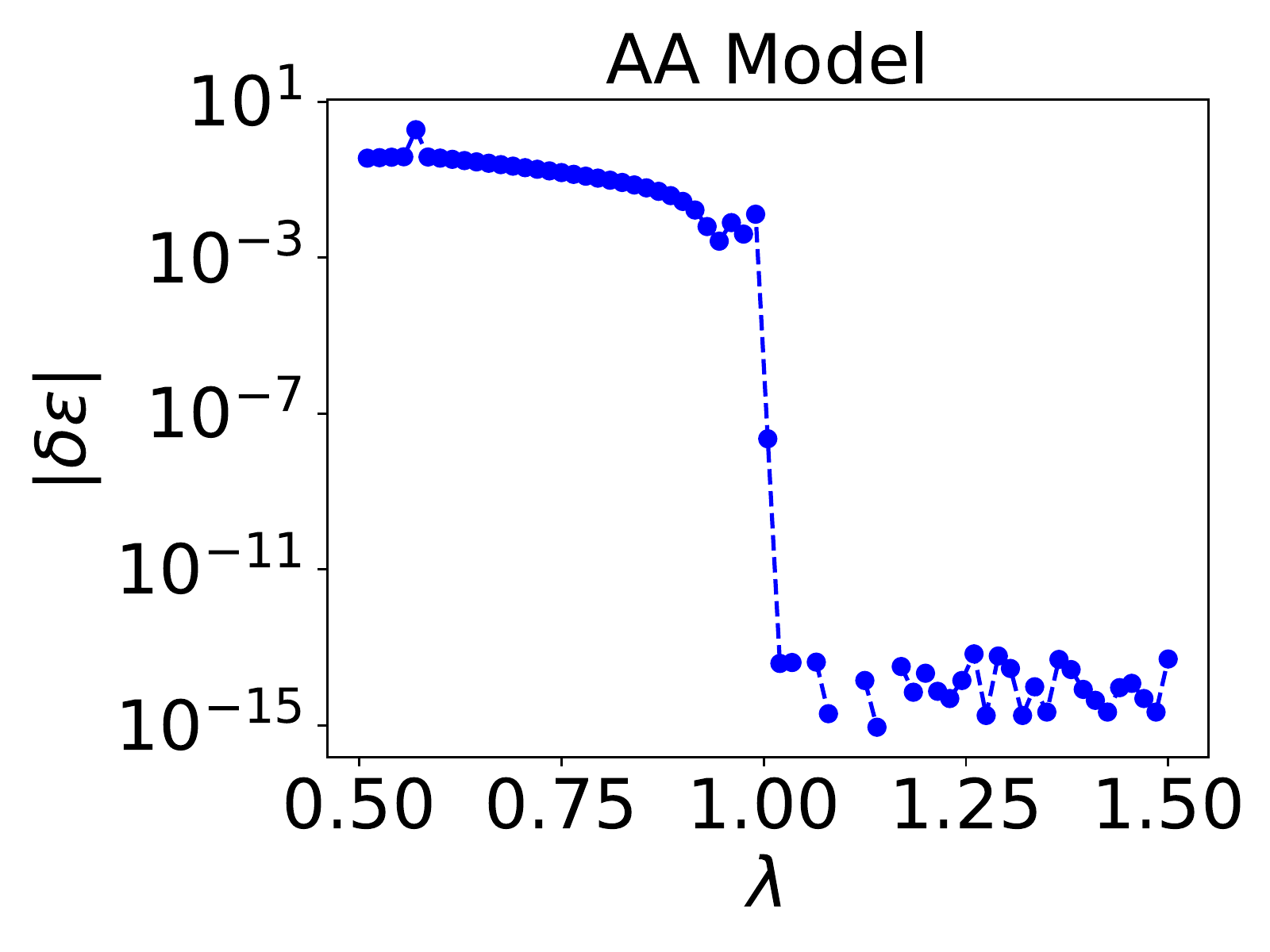}}
    \def\little{\includegraphics[trim={10pt 0pt 10pt 10pt},clip,width=0.14\textwidth]{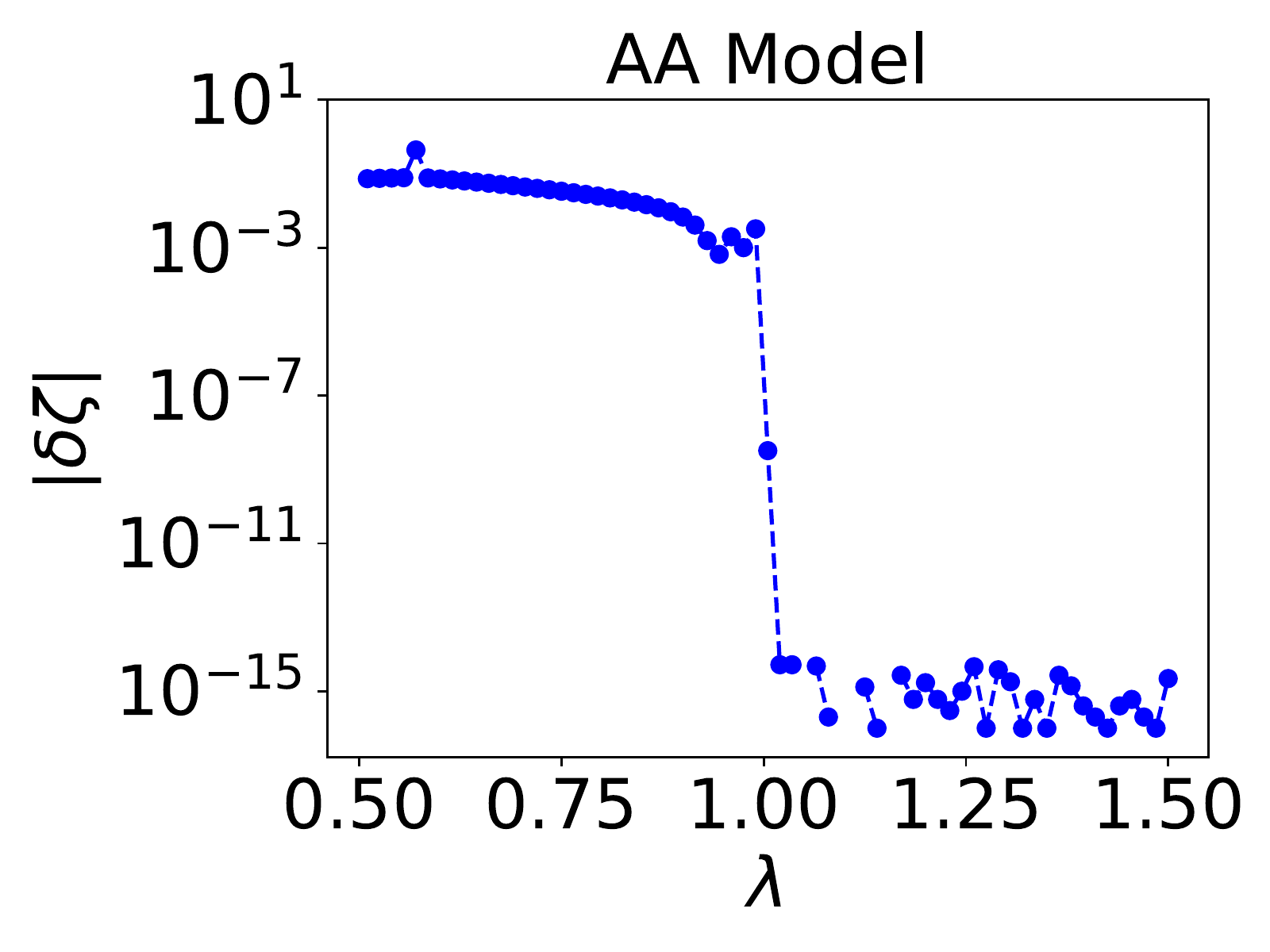}}
    \stackinset{r}{7pt}{t}{15pt}{\little}{\big}
  \end{subfigure}
  \caption{Left panel: disorder average of magnitude of shift in the
    smallest magnitude entanglement energy $<|\delta \epsilon|>$ when
    we change PBC to APBC for RD model as function of $\phi_b$. States
    are delocalized when $\phi_b<2$ and are localized when
    $\phi_b>2$. System size $N=1000$. Disorder average is over $1000$
    samples. Standard deviation of $|\delta \epsilon|$ is plotted in
    lower inset plot. Upper inset plot is the corresponding change in
    $\zeta$. Right panel: magnitude of Shift in the smallest magnitude
    entanglement energy $|\delta \epsilon|$ when we change PBC to APBC
    for AA model as function of $\lambda$. States are delocalized when
    $\lambda<1$ and are localized when $\lambda>1$. Inset plot is the
    corresponding change in
    $\zeta$. $N=2000$. \label{fig:depsilon_RDM_AA}}
\end{figure*}

Now, we focus on the smallest magnitude entanglement energy, the
$\epsilon$ which is closest to zero and has the most contribution
to the EE, Eq. (\ref{EE}). We change the boundary condition from
PBC to APBC and measure the magnitude of shift in the lowest magnitude
entanglement energy $|\delta \epsilon|$ (and the corresponding
$|\delta \zeta|$). For the RD model we plot this shift as a function
of $\phi_b$ in Fig. \ref{fig:depsilon_RDM_AA}. Whereas this shift in
the delocalized phase (i.e. when $\phi_b<2$) is large, it is zero
in the localized phase (it is on the order of $10^{-13}$ for the
chosen system size). Also for the AA model $|\delta \epsilon|$ (and
corresponding $|\delta \zeta|$) is plotted in Fig.
\ref{fig:depsilon_RDM_AA}. The same behavior of $|\delta \epsilon|$
is seen in this model as well.

For both models, we see the shift in the smallest magnitude of the
entanglement energy, sharply determines the phase transition
point. In the delocalized phase $|\delta \epsilon|$ is non-zero and
at the transition pint to localized phase it sharply goes to zero.
Calculation of the $|\delta \epsilon|$ to determine the phase
transition point is numerically more economical rather than
calculation of the $|\delta EE|$ -- where we have to obtain the
entire spectrum -- specially that there are numerical packages
(like ARPACK) by which we can obtain the smallest eigen-value
efficiently.

\begin{figure*}
  \centering
  \begin{subfigure}{}%
    \includegraphics[width=0.31\textwidth]{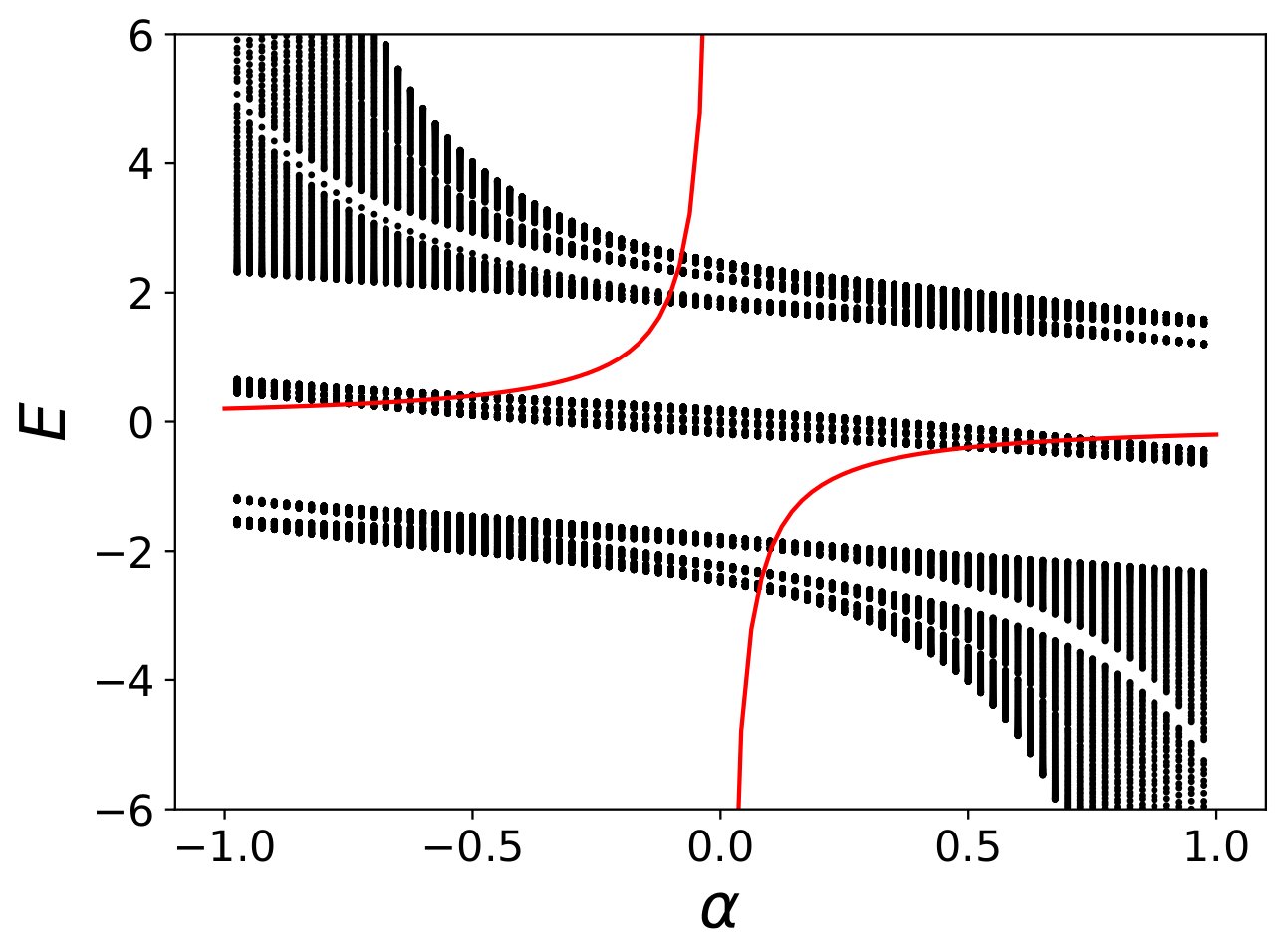}
  \end{subfigure}%
  ~%
  \begin{subfigure}{}%
    \includegraphics[width=0.32\textwidth]{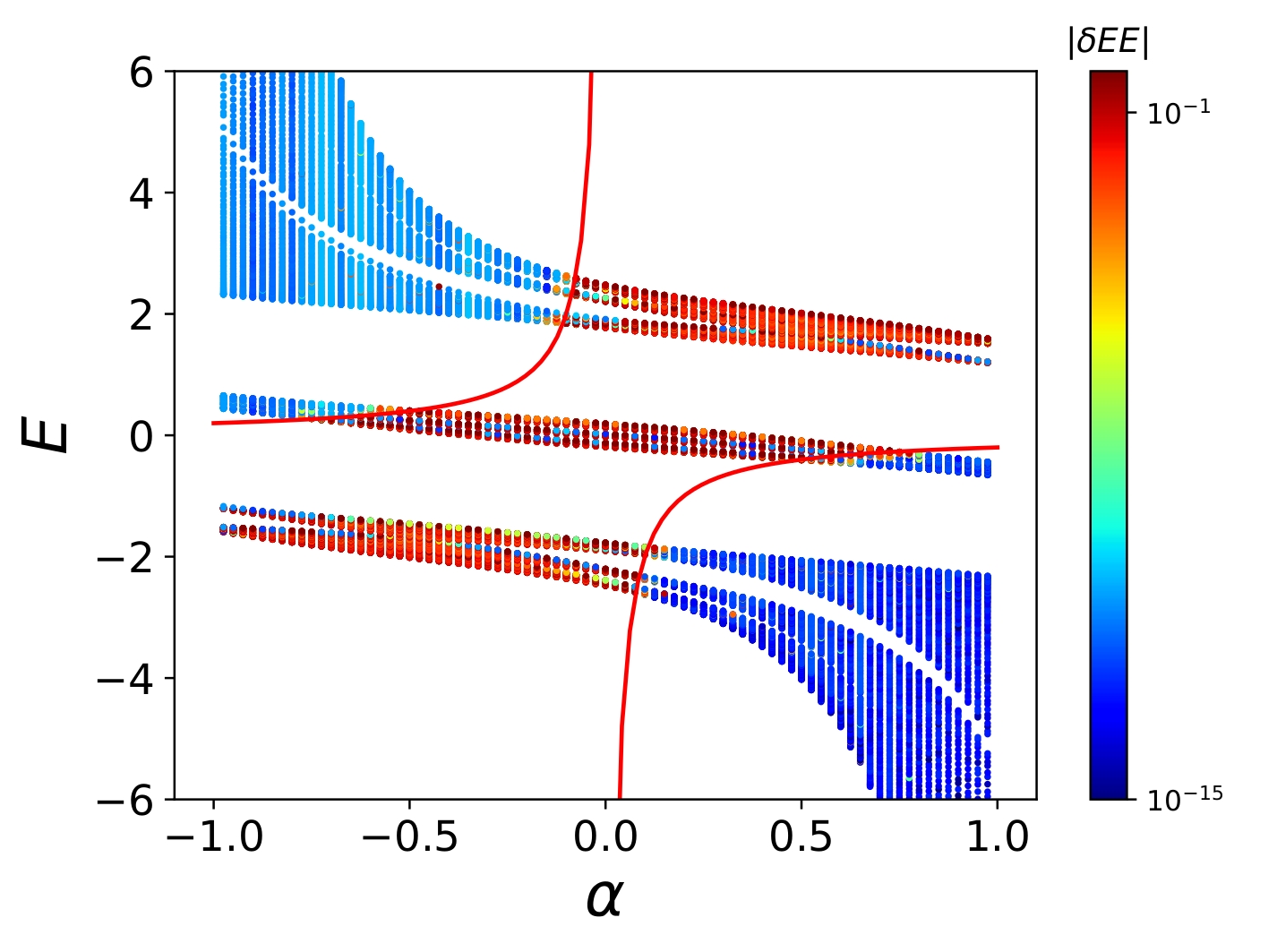}
  \end{subfigure}
  ~%
  \begin{subfigure}{}%
    \includegraphics[width=0.32\textwidth]{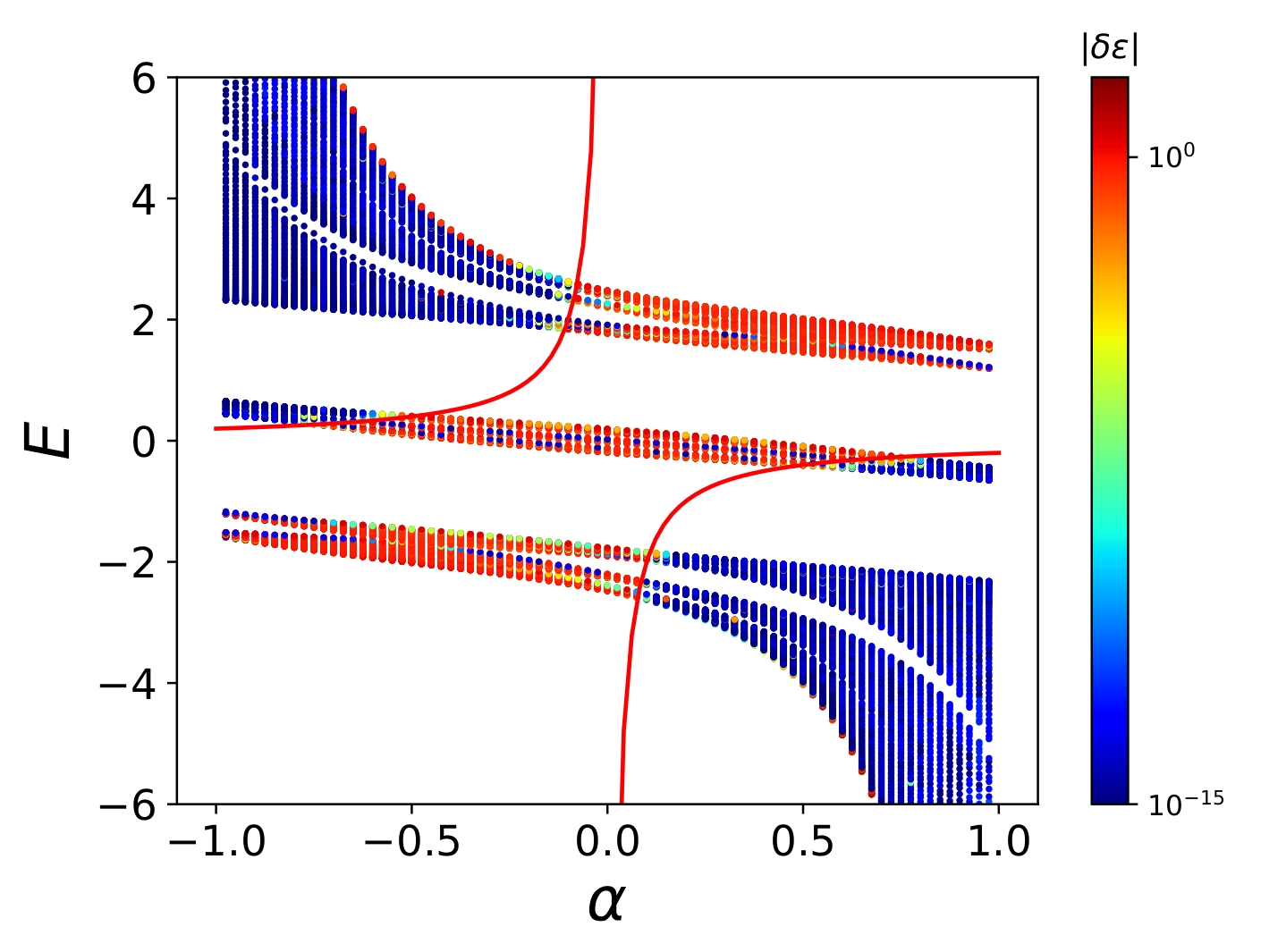}
  \end{subfigure}%
  \caption{(color online) Left panel: spectrum of the Hamiltonian of
    Eq. (\ref{oh}) with the on-site energies of Eq. (\ref{gAA}). The
    mobility edges which are determined by Eq. (\ref{me}) are plotted
    in red. Middle panel: magnitude of shift in the entanglement entropy
    $|\delta EE|$ when boundary condition changes from PBC to APBC for
    the whole spectrum and for $\alpha$ between $-1$ and $1$. Right
    panel: magnitude of Shift in the smallest magnitude entanglement energy
    $|\delta \epsilon|$. $N=500$. Colorbars are plotted in log
    scale. \label{fig:d_gAA}}
\end{figure*}

Next, we consider a model that has mobility edges (contrary to RD
and AA models we have considered up to now). Namely, we consider
Hamiltonian of Eq. (\ref{oh}) with the on-site energies of Eq.
(\ref{gAA}). Mobility edges are determined by the Eq. (\ref{me}),
and we set $t=-1$. We calculate the shift in the entanglement
energies to see how good this shifts can locate the mobility
edges. The results are plotted in Fig. \ref{fig:d_gAA}.  We go
though $\alpha$ from $-1$ to $1$ and calculate the eigen-energy
spectrum at each point. For the allowed eigen-energies we
calculate the change in the entanglement entropy $|\delta EE|$, and
the change in the smallest magnitude entanglement energy $|\delta
\epsilon|$. The mobility edge between extended and localized states
can be located by $|\delta EE|$ and $|\delta \epsilon|$ fairly well.
This provides additional evidence for our conjecture that $|\delta
EE|$ and/or $|\delta \epsilon|$ can provide us with important
information about localization properties of a given system.

\section{Concluding Remarks}\label{conclusions}
We examined the effect of the change in the boundary conditions on the
entanglement properties of the system. Namely, we changed the boundary
conditions from PBC to APBC and studied the change in the spectrum of
the entanglement Hamiltonian and also in the entanglement entropy.
By using one-dimensional free fermion models which have
LD phase transition, we showed numerically that in
the delocalized phase the spectrum of the entanglement Hamiltonian and
thus entanglement entropy changes, but in the localized phase the
shift is negligible. We also studied the shift in one of the
eigen-values of the entanglement Hamiltonian, the smallest magnitude
entanglement energy, and we showed that this shift is enough to determine the
phase transition point: shift $|\delta \epsilon|$ is non-zero in the
delocalized phase and sharply goes to zero in localized phase. Thus we
verified that the shift in the entanglement Hamiltonian spectrum can
be identified as a \emph{new} phase detection parameter.

We studied the LD phase transition by examining the ground state
entanglement Hamiltonian instead of original Hamiltonian of the
system. The next question would be: can we obtain the conductance
properties of the system by examining the entanglement Hamiltonian
rather than Hamiltonian of the system? In addition, $\delta
\epsilon$ as a phase detection parameter, deserves more studies
for models with randomness. For example it is interesting to study
$\delta \epsilon$ and its distribution in two and
three-dimensional Anderson models with weak localization and
localization-delocalization phase transitions. Clearly, such
issues would be of interest and we intend to address these in the
future.

\acknowledgements This research is supported by Shiraz University
research council and National Merit Foundation of Iran.


\begin{references}

\bibitem{ref:Horodecki}
  R. Horodecki, P. Horodecki, M. Horodecki, and K. Horodecki,
  Rev. Mod. Phys. \textbf{81}, 865 (2009).

\bibitem{ref:chuang}
  M. A. Nielsen, I. L. Chuang, Quantum Computation and Quantum
  Information (Cambridge University Press) (Cambridge) (2010).

\bibitem{ref:Ekert}
  A. Ekert, Phys. Rev. Lett \textbf{67}, 661 (1991).

\bibitem{ref:Steane}
  A. Steane, Rep. Prog. Phys. \textbf{61}, 117 (1998).

\bibitem{ref:Gisin}
  N. Gisin, G. Ribordy, W. Tittel, and H. Zbinden,
  Rev. Mod. Phys. \textbf{74}, 145 (2002).

\bibitem{ref:osterloh}
  A. Osterloh, L. Amico, G. Falci, R. Fazio, Nature \textbf{416},
  608-610 (2002).

\bibitem{ref:amico}
  L. Amico, R. Fazio, A. Osterloh, and V. Vedral,
  Rev. Mod. Phys. \textbf{80}, 517 (2008).

\bibitem{ref:haque}
  M. Haque, O. Zozulya, and K. Schoutens,
  Phys. Rev. Lett. \textbf{98}, 060401 (2007).

\bibitem{ref:kitaev}
  A. Kitaev and J. Preskill, Phys. Rev. Lett. \textbf{96}, 110404 (2006).


\bibitem{ref:chandran}
  A. Chandran, V. Khemani, and S. L. Sondhi,
  Phys. Rev. Lett. \textbf{113}, 060501 (2014).

\bibitem{ref:min-fongyang}
  Min-Fong Yang, Phys. Rev. A \textbf{71}, 030302(R) (2005).

\bibitem{ref:vedral}
  V. Vedral, M. B. Plenio, M. A. Rippin, and P. L. Knight,
  Phys. Rev. Lett. \textbf{78}, 2275 (1997).

\bibitem{ref:vidal}
  G. Vidal, J. I. Latorre, E. Rico, and A. Kitaev,
  Phys. Rev. Lett. \textbf{90}, 227902 (2003).

\bibitem{ref:osborne}
  T. J. Osborne and M. A. Nielsen, Phys. Rev. A \textbf{66},
  032110 (2002).

\bibitem{ref:shi}
  S. J.  Gu, S. S.  Deng, Y. Q. Li, and H. Q. Lin,
  Phys. Rev. Lett. \textbf{93}, 086402 (2004).

\bibitem{ref:vidal2}
  J. Vidal, G. Palacios, and R. Mosseri, Phys. Rev. A
  \textbf{69}, 022107 (2004).


\bibitem{ref:mondragon}
I. Mondragon-Shem, M. Khan, and T. L. Hughes,
Phys. Rev. Lett. \textbf{110}, 046806 (2013).

\bibitem{ref:andrade}
E. C. Andrade, M. Steudtner, M. Vojta, J. Stat. Mech. P07022 (2014)
.


\bibitem{ref:mondragon2}
I. Mondragon-Shem and T. L. Hughes,
Phys. Rev. B \textbf{90}, 104204 (2014).

\bibitem{ref:othercut}
M. Legner and T.  Neupert,
Phys. Rev. B \textbf{88}, 115114 (2013). O. S. Zozulya, M. Haque,
K. Schoutens, and E. H. Rezayi, Phys. Rev. B \textbf{76}, 125310
(2007). M. C. Arnesen, S. Bose, and V. Vedral,
Phys. Rev. Lett.\textbf{87}, 017901 (2001). Ronny Thomale,
D. P. Arovas, and B. A. Bernevig, Phys. Rev. Lett. \textbf{105},
116805 ( 2010).


\bibitem{ref:wei}
T. C. Wei, D. Das, S. Mukhopadyay, S. Vishveshwara, and P. M. Goldbart
Phys. Rev. A \textbf{71} 060305(R) (2005).

\bibitem{ref:love} P. J. Love, A.M. van den Brink, A.Y. Smirnov, et
al. Quantum Inf Process  6:187 (2007).

\bibitem{ref:chhajlany}
R. W. Chhajlany, P. Tomczak, A. Wojcik, and J. Richter
Phys. Rev. A \textbf{75}, 032340 (2007).


\bibitem{ref:afshin1}
A. Montakhab and A. Asadian, Phys. Rev. A \textbf{77}, 062322 (2008).

\bibitem{ref:afshin2}
A. Montakhab and A. Asadian, Phys. Rev. A \textbf{82}, 062313 (2010).


\bibitem{ref:LiHaldane}
  H. Li and F. D. M. Haldane, Phys. Rev. Lett. \textbf{101}, 010504
  (2008).

\bibitem{ref:pouranvariyang1}
  M. Pouranvari, K. Yang, Phys. Rev. B \textbf{88}, 075123 (2013).

\bibitem{ref:pouranvariyang2}
  M. Pouranvari, K. Yang, Phys. Rev. B \textbf{89}, 115104 (2014).

\bibitem{ref:pouranvariyang5}
  M. Pouranvari, K. Yang, Phys. Rev. B \textbf{92}, 245134 (2015).








\bibitem{ref:berkovitsprl2015}
R. Berkovits,  Phys. Rev. Lett. \textbf{115}, 206401 (2015).

\bibitem{ref:chakravarty}
  S. Chakravarty, Int. J. Mod. Phys. B \textbf{24}, 1823 (2010).

\bibitem{ref:zhao}
  A. Zhao, R. L. Chu, S. Q.  Shen,  Phys. Rev. B \textbf{87}, 205140 (2013).


\bibitem{ref:pouranvariyang4} M. Pouranvari, Y. Zhang, K. Yang,
Advances in Condensed Matter Physics, vol. 2015, 397630 (2015).

\bibitem{ref:berkovitsprl2012}
R. Berkovits,  Phys. Rev. Lett. \textbf{108}, 176803 (2012).


\bibitem{ref:bardarson}
  J. H. Bardarson, F. Pollmann, and J. E. Moore,
Phys. Rev. Lett. \textbf{109}, 017202 (2012).

\bibitem{ref:nag}
  S. Nag, A. Garg, arXiv:1701.00236.

\bibitem{ref:filho}
  J. L. C. da C. Filho, A. Saguia, L. F. Santos, M. S. Sarandy,
  arXiv:1705.01957.

\bibitem{ref:geraedts}
  S. D. Geraedts, N. Regnault, R. M. Nandkishore,
  arXiv:1705.00631.

\bibitem{ref:khemani}
  V. Khemani, S. P. Lim, D. N. Sheng, and D. A. Huse,
Phys. Rev. X \textbf{7}, 021013 (2017).

\bibitem{ref:yang}
  Z. Yang, A. Hamma, S. M. Giampaolo, E. R. Mucciolo, C. Chamon,
  arXiv:1703.03420.

\bibitem{ref:berkovits}
  R. Berkovits, Ann. Phys. (Berlin), 1700042 (2017).

\bibitem{ref:vidmar}
  L. Vidmar, L. Hackl, E. Bianchi,  and M. Rigol, arXiv:1703.02979.



\bibitem{ref:laflorencie2006}
N. Laflorencie, E. S. Sorensen, M. S. Chang, and I. Affleck,
Phys. Rev. Lett.\textbf{96}, 100603 (2006).


\bibitem{ref:levine}
G. C. Levine, Phys. Rev. Lett. \textbf{93}, 266402 (2004).

\bibitem{ref:peschel2005}
I. Peschel, J. Phys. A: Math. Gen. \textbf{38} 4327 (2005).


\bibitem{ref:Shklovskii}
  B. I. Shklovskii, B. Shapiro, B. R. Sears, P. Lambrianides, and
  H. B. Shore, Phys. Rev. B \textbf{47}, 11487 (1993).

\bibitem{ref:edward} J. T. Edwards and D. J. Thouless, J. Phys. C:
Solid State Phys., Vol. 5 (1972).


\bibitem{ref:anderson}
  P.W. Anderson, D. J. Thouless, E. Abrahams, D. S. Fisher,
  Phys. Rev. B {\bf 22}, 3519, (1980).

\bibitem{ref:economou}
  E. N. Economou and C. M. Soukoulis, Phys. Rev. Lett. \textbf{46}, 618 (1981).

\bibitem{ref:kramer}
  B. Kramer, A. MacKinnon, Rep. Prog. Phys. \textbf{56} 1469-1564 (1993).

\bibitem{ref:footenote}
  for a description of the $\ket{MEM}$ look at
  Ref. [\onlinecite{ref:pouranvariyang2}]

\bibitem{ref:diener}
  R. B. Diener, G. A. Georgakis, J. Zhong, M. Raizen, and Q. Niu,
  Phys. Rev. A {\bf 64}, 033416 (2001).

\bibitem{ref:lahini}
  Y. Lahini, R. Pugatch, F. Pozzi, M. Sorel,
  R. Morandotti, N. Davidson, and Y. Silberberg,
  Phys. Rev. Lett. \textbf{103}, 013901 (2009).




\bibitem{correl}
  I. Peschel, J. Phys. A: Math. Gen. \textbf{36}, L205 (2003).

\bibitem{dunalp}
  D. H. Dunlap, H-L. Wu, and P. W. Phillips, Phys. Rev. Lett {\bf 65},
  88 (1990).

\bibitem{ref:dassarma}
  Sriram Ganeshan, J. H. Pixley, and S. Das Sarma,
  Phys. Rev. L. \textbf{114}, 146601 (2015)


\bibitem{ref:aubryandre} S. Aubry, G. André, Ann. Israel. Phys.
Soc. 3, 133 (1980).
\end{references}
\end{document}